\documentclass[12pt]{article}
\usepackage[top=1in, bottom=1in, left=1in, right=1in]{geometry}
\usepackage[english]{babel}
\usepackage{atbegshi,cite}
\usepackage{amsmath,amssymb,amsbsy,amstext, amsthm, simplewick}
\usepackage{graphicx}
\usepackage{amsfonts}
\usepackage[small]{caption}
\usepackage{upgreek}
\usepackage[titletoc]{appendix}
\usepackage{setspace}
\usepackage{comment}
\usepackage[usenames,dvipsnames,table]{xcolor}
\usepackage[colorlinks=true,urlcolor=blue
,anchorcolor=blue,citecolor=blue,filecolor=blue,linkcolor=blue,menucolor=blue
]{hyperref}

\newcommand{\newc}{\newcommand}
\newc{\fpi}{f_{\pi}}
\newc{\etap}{\eta^{\prime}}
\newc{\llll}{\langle\lambda\lambda\rangle}
\newc{\FFd}{F^a\tilde F^a}
\newc{\qbar}{{\overline q}}
\newc{\TR}{{\rm Tr}}
\newc{\Kahler}{K\"ahler }
\newc{\Zbb}{{\mathbb Z}}
\newc{\Rt}{{{\mathbb R}^3}}
\newc{\Rf}{{{\mathbb R}^4}}
\newc{\Sth}{{{\mathbb S}^3}}
\newc{\SthSo}{{{\mathbb S}^3\times{\mathbb S}^1}}
\newc{\Stw}{{{\mathbb S}^2}}
\newc{\StwSo}{{{\mathbb S}^2\times{\mathbb S}^1}}
\newc{\So}{{{\mathbb S}^1}}
\newc{\zt}{{{\mathbb Z}_2}}
\newc{\RtSo}{{{\mathbb R}^3\times{\mathbb S}^1}}
\newc{\RfSo}{{{\mathbb R}^4\times{\mathbb S}^1}}
\newc{\scriminus}{{\cal I}^-}
\newc{\scriplus}{{\cal I}^+}
\newc{\mpl}{M_p}
\newc{\Ricci}{\mathcal{R}}
\newc{\bv}{\phi}
\newc{\calU}{{\cal U}}
\newc{\calK}{K}
\newc{\calUi}{{\cal U}^{-1}}
\newc{\calG}{{\cal G}}
\newc{\calI}{{\cal I}}
\newc{\calO}{{\cal O}}
\newc{\calQ}{{\cal Q}}
\newc{\calOb}{{\cal O}^\dagger}
\newc{\hphi}{{\hat\phi}}

\theoremstyle{plain}
\theoremstyle{plain} 
\theoremstyle{plain} 
\theoremstyle{plain}
\theoremstyle{plain}
\theoremstyle{plain}

\renewcommand{\title}[1]{{\Large\bf\flushleft{#1}}\vspace*{3ex}\\}
\renewcommand{\author}[2]{{\noindent\hspace*{2.5em}\large#1}
                     \footnote{Electronic mail: $\mathtt{#2}$}\\}

\renewcommand\vec{\mathbf}
\newcommand{\Mpl}{M_p}
\newcommand{\beq}{\begin{equation}}
\newcommand{\eeq}{\end{equation}}
\newcommand{\Leff}{L_\mathrm{eff}}
\newcommand{\Lmin}{L_\mathrm{min}}

\begin{document}
\begin{titlepage}
\begin{flushright}
{ 
FERMILAB-PUB-21-315-T\\
}
\end{flushright}

\vskip 2.2cm

\begin{center}

{\large \bf Densities of States and the CKN Bound }
\vskip 1.4cm

{ Nikita Blinov$^{(a,\, b)}$ and Patrick Draper$^{(c)}$ }
\\
\vskip 1cm
{$^{(a)}$ Fermi National Accelerator Laboratory, Batavia, IL, 60510, USA}\\
{$^{(b)}$ Kavli Institute for Cosmological Physics, University of Chicago, Chicago, IL USA}\\
{$^{(c)}$ Illinois Center for Advanced Studies of the Universe \&\\Department of Physics, University of Illinois, Urbana, IL 61801}\\
\vspace{0.3cm}
\vskip 4pt

\vskip 1.5cm

\begin{abstract}
The holographic principle implies that quantum field theory overcounts the number of independent degrees of freedom in quantum gravity. 
An argument due to Cohen, Kaplan, and Nelson (CKN) suggests that the number of degrees of freedom well-described by QFT is even smaller than required by holographic bounds, and CKN interpreted this result as indicative of a correlation between the UV and IR cutoffs on QFT.  
Here we consider an alternative interpretation in which the QFT degrees of freedom are depleted as a function of
scale. We use a simple recipe to estimate the impact of depleted densities of states on precision observables, including the Lamb shift and lepton $g-2$. Although these observables are not sensitive to the level of depletion motivated by gravitational considerations, the phenomenological exercises also provide an interesting test of  quantum field theory that is independent of underlying quantum gravity assumptions. 
A depleted density of states can also render the QFT vacuum energy UV-insensitive,  reconciling the success of QFT in describing ordinary particle physics processes
 and its apparent failure in predicting the cosmological constant.
\end{abstract}

\end{center}

\vskip 1.0 cm

\end{titlepage}
\setcounter{footnote}{0} 
\setcounter{page}{1}
\setcounter{section}{0} \setcounter{subsection}{0}
\setcounter{subsubsection}{0}
\setcounter{figure}{0}

%\doublespacing
\onehalfspacing

%\tableofcontents
\section{Introduction}
\label{sec:intro}
We do not know how realistic quantum field theories emerge as a low energy approximation to a complete theory of quantum gravity with a positive cosmological constant. 
However, it is likely that QFT contains too many degrees of freedom, even with an ultraviolet cutoff $\Lambda$ well below the Planck scale. In QFT, the maximum entropy localized in a box of size $L$ scales extensively as  $S\sim (\Lambda  L)^3$, while the holographic principle limits the number of degrees of freedom in quantum gravity to $S\sim M_p^2L^2$. Furthermore, this overabundance of QFT states does not mean that most of the states in quantum gravity have to behave like bulk QFT states. A hot box containing QFT degrees of freedom collapses to a black hole when 
$T^4 L^3/\mpl^2 \sim L$, corresponding to an entropy $S\sim (\mpl L)^{3/2} \ll \mpl^2 L^2$. 
It could well be that the number of degrees of freedom in the correct quantum theory of gravity that are well-described by bulk quantum field theory is only of order $L^{3/2}$~\cite{Cohen:1998zx,Banks:2019oiz}.

Let us first recall some old ideas for how bulk particles might realize holographic bounds on degrees of freedom~\cite{Thorn:1991fv,tHooft:1993dmi,Susskind:1993aa,Susskind:1994vu}. In~\cite{Susskind:1994vu} particles were modeled as a collection of indivisible partons in light front quantization. The number of partons one should ascribe to a particle grows with the lab frame energy $\epsilon$ and the resolution with which its momentum can be measured: $N_{parton} \sim \epsilon L$, where we describe the momentum resolution as an infrared length scale, $\Delta p= 1/L$. Once the number of partons exceeds some inverse coupling of the microscopic quantum gravity theory, for example the string coupling $g_s^2$, the size of the particle $R$ begins to grow rapidly with each additional parton. It scales as $M_p R \sim (\epsilon L)^{1/(d-1)}$ in $d$ spatial dimensions, so that $M_p R\sim \sqrt{\epsilon L}$ in $d=3$.

When the particle size is so large that it can no longer be localized on length scales of order the inverse energy, it ceases to be point-like  and cannot be regarded as an excitation of one independent degree of freedom. However, in the parton model this limit depends on the momentum resolution.  
We could interpret the saturation of $R=1/\epsilon$ as supplying a bound on the typical momentum spacing between individual particle degrees of freedom around $\epsilon$,
\begin{align}
1/\Lmin = \epsilon^{3}/\mpl^2.
\label{eq:Lminholo}
\end{align}
The spacing grows with energy, rapidly depleting the single particle density of states (DOS) $g(\epsilon)$. 
In a box of fixed size $L$, there is an energy scale $\hat\epsilon=(\mpl^2 /L)^{1/3}$ at which the momentum resolution  transitions from $1/L$ to $1/L_{min}$.  Here the relativistic DOS  transitions from  ordinary QFT scaling $g\sim L^3\epsilon^2$ to $g\sim\Lmin^3\epsilon^2$. Most of the degrees of freedom live around $\hat\epsilon$, and the maximum entropy is of order $(L\hat\epsilon)^3\sim (L \mpl)^2$,  satisfying the holographic bound. Alternatively, the same result is obtained by counting states according to the scaling $g\sim \partial_\epsilon(L^3/R^3)$ for $R>1/\epsilon$.

We see that the partonic model realizes the holographic principle by a scale-dependent depletion of the independent degrees of freedom, relative to the density of states of an ordinary bulk field theory in a fixed volume. 

However, as alluded to above, there is an argument due to Cohen, Kaplan, and Nelson (CKN) that the number of degrees of freedom that can be well-described by bulk QFT might be even smaller than what is implied by holography~\cite{Cohen:1998zx}. CKN interpreted the entropy scaling $S\sim (M_p L)^{3/2}$  as implying a correlation between the UV and IR cutoffs $\Lambda$ and $L$ on QFT: requiring $(\Lambda L)^3  \lesssim ( \mpl L)^{3/2}$, one arrives at the bound 
\begin{align}
L \lesssim M_p/\Lambda^2.
\label{eq:CKNrelation}
\end{align}
For a weak scale UV cutoff, $L$ is the order of 1 cm when the bound is saturated. What this IR scale means, however, is unclear. In Ref.~\cite{Cohen:1998zx} it was interpreted as an IR momentum cutoff on Feynman graphs for a given UV cutoff, and recently there has been renewed interest in this interpretation of the bound and its possible implications for precision measurements~\cite{Bramante:2019uub, Cohen:2021zzr, Davoudiasl:2021aih}. In fact this interpretation suggests that the effects of quantum gravity on precision observables are so large that they may be detectable following plausible experimental improvements~\cite{Cohen:1998zx,Bramante:2019uub, Cohen:2021zzr, Davoudiasl:2021aih}.

In this paper we consider a different interpretation, introduced in~\cite{Banks:2019arz}, where $1/L$ is taken to be  a  bound on a characteristic momentum spacing between independent degrees of freedom that can be well-described by bulk quantum field theory. In other words, $L$ controls a depletion of the single-particle density of  states of ordinary QFT, and the depletion may be stronger than what is required by holography alone.

An insightful observation made in~\cite{Banks:1995uh,Cohen:1998zx} was phrased in~\cite{Cohen:1998zx} as follows: ``There is in
fact no evidence that fields at present experimental energies can fluctuate independently over a region as large as our
horizon." Over how large a region {\emph{can}} localized excitations of a typical energy fluctuate independently? This is both an empirical question and a question of principle.  In order to localize an excitation  of energy $\sim \epsilon$ in, say, a minimal region of size $1/\epsilon$, inside of a much larger  region $\sim L$, there must be many independent modes around $\epsilon$ with momentum spacing $\Delta p\sim 1/L$.  Since no real experiment is sensitive to arbitrarily small differences in momenta or energy, however, there are only empirical lower bounds on the QFT single-particle density of states. As a matter of principle, we interpret the bound~(\ref{eq:CKNrelation}) as a fundamental \emph{upper} limit on the DOS. 
The bound~(\ref{eq:CKNrelation}) suggests  that the relativistic DOS is depleted above some energy scale, behaving as
\beq
g\sim \begin{cases} L^3 \epsilon^2 ~& \epsilon^2 \ll \mpl/L\\
\mpl^3/\epsilon^4 & \epsilon^2 \gg \mpl/L
\end{cases}.
\label{eq:naiveDDOS}
\eeq
We can parametrize the scale-dependent depletion in terms of an effective box size at different energy scales, $\Leff(\epsilon)=\min ( L, \mpl/\epsilon^2)$.  This interpretation is similar to the holographic partonic picture described above, but with different scalings, such that $\Leff$ of Eq.~(\ref{eq:naiveDDOS}) is smaller than the holographic $\Lmin$ of Eq.~(\ref{eq:Lminholo}).

In this work we will consider first the empirical question: how well has the single particle density of states been tested at various energy scales? This assessment is independent of any more fundamental motivations about quantum gravity, and represents an interesting, somewhat unconventional  axis along which quantum field theory can be tested. Simple estimates can be made based on the sensitivity of detectors, while less trivial estimates arise from precision measurements of observables that involve sums over intermediate states at characteristic energies $\epsilon$. In the latter category we consider atomic spectra and leptonic $g-2$, using finite volume techniques to impose a discretization of the state space. In this way we can associate an empirical lower bound on the IR scale $\Leff(\epsilon)$ from these measurements. We  compare the empirical bounds on $\Leff$ to (\ref{eq:naiveDDOS}), finding that current precision falls well short of the modifications suggested by these gravitational arguments. Thus this is an interpretation of the CKN bound that does minimal violence to the predictions of ordinary QFT.

We then reexamine the CKN ``hot box" thought experiments using the depleted DOS in Eq.~\eqref{eq:naiveDDOS} and generalize it to other power laws. We note that there are some nonthermal, high occupancy states which would collapse to a black hole in a gravitational theory and which are not removed either by depleting the single-particle DOS or by placing correlated cutoffs on Feynman integrals.  Some states must be excised by other effects.

We emphasize that we do not have any concrete proposal for how a fundamental depletion of QFT modes should be implemented, nor do we know how Lorentz invariance and locality should emerge. In Matrix Theory, a theory which realizes the holographic principle yet gives rise to a supergravity theory at long distances, Lorentz invariance and locality  of scattering amplitudes are thought to result from delicate cancellations and depend intricately on the BPS nature of the partons~\cite{Banks:1996vh}. We will at least not assume any modification to the relativistic  dispersion relations, which provide the most sensitive experimental probes of Lorentz invariance, and our results for $g-2$ and the Lamb shift differ from the predictions of ordinary relativistic QFT by  ${\mathcal{O}}(1/\mpl^2)$.

These exercises have curious implications for the cosmological constant problem. The QFT contribution to the total vacuum energy can be written as
\begin{align}
\Delta \rho \sim V^{-1}\int \tilde g (\epsilon) \epsilon d\epsilon
\end{align}
which is quartically divergent for the usual DOS. This expression depends on the graded DOS $\tilde g = (-1)^{F}g$ where $F=0$ (1) for bosons (fermions). Exact supersymmetry solves the cosmological constant problem within quantum field theory alone because it depletes the graded DOS completely,  leaving unaffected the ordinary DOS for different species. A few other distinct and fascinating examples in quantum field theory also yield cancellations in the graded DOS~\cite{Dienes:1994np,Cherman:2018mya}. We are instead considering the possibility that the DOS is depleted for each species separately, and that this depletion is intrinsically  gravitational -- it disappears in the $M_p\rightarrow\infty$ limit. With the depletion~(\ref{eq:naiveDDOS}), for example, the fine-tuning problem is removed. Moreover, we will conclude that the bounds on the DOS obtained from precision particle physics measurements are far from definitively establishing that the cc problem is a problem, without invoking a substantial extrapolation.

\section{Empirical Probes of the QFT Density of States}
\label{sec:depletion_from_finite_volume}

The density of states  enters into the sum over intermediate states in precision processes. For such processes we can estimate the sensitivity to the DOS using finite-volume techniques. Placing a theory in a periodic box of volume $L^3$,  the DOS of a massless particle is depleted in a scale-independent way from the continuum  to $g\sim \epsilon^2 L^3$. Standard techniques can be used to compute the $L$-dependent effects on the process, typically considered as a power series in $1/L$ for large boxes. Since we are interested in the sensitivity of observables to the DOS at different scales, we proceed as follows.
 Given an observable $\calO$ computed in perturbation theory, we  isolate in some manner the perturbative corrections to $\calO$ from  intermediate states where the individual particles all have some characteristic energy $\epsilon_*$. Call these contributions $\calO_{\epsilon_*}$. Placing the theory in a box of size $L$, we  then compute the $1/L$ corrections to $\calO_{\epsilon_*}$, denoted $\calO^L_{\epsilon_*}$. Comparing $\calO^L_{\epsilon_*}$ to the experimental precision with which $\calO$ is known, we can place a lower bound on the box size, $L>L_\calO$. We identify this bound as the current experimental limit on the depleted density of states at $\epsilon_*$, parametrized by the effective length scale $\Leff$: 
\begin{align}
\Leff(\epsilon_*)>L_\calO.
\end{align}

In other words, the impact of the depletion can be approximated as a finite-volume effect in a box of size $\Leff(\epsilon_*)$.  Corrections 
to this estimate depend on $\Delta \epsilon / \epsilon_*$, where $\Delta \epsilon$ is the range of energies allowed to contribute to $\calO_{\epsilon_*}$.

Such estimates can be performed entirely numerically. However, we will find it convenient to take a slightly different approach. We will compute the entire $L$-dependent effects on observables $\calO$ analytically as a series in $1/L$, then inspect the terms to see which intermediate energy scales  contribute dominantly to each power of $1/L$. 

What these methods lack in sophistication, we hope they make up for in generality: since at best we obtain only conservative, order-of-magnitude bounds on $\Leff$, they might have some resilience against whatever detailed form the  actual depletion takes.

We begin by computing the modification to the Coulomb potential and infer bounds on $\Leff(\epsilon_\star=\alpha m_e)$ from atomic physics. We then consider state depletion effects on $\Leff(\epsilon_\star=m_e)$ from $g-2$, since that precisely-known quantity has been of particular interest in the CKN-related literature~\cite{Cohen:1998zx,Banks:2019arz, Bramante:2019uub, Cohen:2021zzr, Davoudiasl:2021aih}.  Finally we briefly consider bounds at higher energies based on the measured resolution of detectors.

\subsection{Coulomb Forces and Bound State Energies}
\label{sec:coulomb}
The depletion of single particle states can affect the electromagnetic force in atomic physics.
In standard QFT the Coulomb potential is 
\beq
V(\vec{r}) =-\int \frac{d^3 q}{(2\pi)^3} \frac{e^2}{\vec{q}^2} e^{i\vec{q}\cdot \vec{r}} 
= -\frac{e^2}{2\pi^2r} \int_0^\infty d\tilde{q}\; \frac{\sin\tilde{q}}{\tilde{q}}.
\label{eq:qed_pot}
\eeq
In old-fashioned perturbation theory, the integral over $q$ corresponds to a sum over 
on-shell intermediate states with momentum $\vec{q}$ and energy $E = |\vec{q}|$.
We leave the final integral over $\tilde{q} = qr$ unevaluated in Eq.~(\ref{eq:qed_pot}) to highlight the fact that 
the largest contribution comes from modes with $qr \lesssim 1$. 

As described above, we  model a depletion of the DOS  
by putting the system in a box of size $L$. 
Following~\cite{Hasenfratz:1989pk,Kronfeld:2002pi} we rewrite the integrand in Eq.~(\ref{eq:qed_pot}) in Schwinger parametrization and discretize 
Cartesian momenta  $\vec{q}_i \to (2\pi n_i)/L$:
\begin{align}
V(\vec{r}) & =  -e^2\int_0^\infty d\rho\frac{1}{L^3}\prod_{i=1}^3\left\{\sum_{n_i=-\infty}^{+\infty}\exp\left(-\rho \left(\frac{2\pi n_i}{L}\right)^2 + i\left(\frac{2\pi n_i}{L}\right)\vec{r}_i\right)\right\} \nonumber\\ 
& = -\frac{e^2}{L^3} \int_0^\infty d\rho \prod_{i=1}^3\theta_3\left(-\frac{\pi \vec{r}_i}{L},e^{-4\pi^2 \rho/L^2}\right),
\label{eq:finite_volume_potential}
\end{align}
where 
\beq
\theta_3(z,\tau) = \sum_{n=-\infty}^{+\infty} \exp(i\pi n^2 \tau + i2\pi n z)
\label{eq:elliptic_theta_def}
\eeq
is the elliptic theta function. 
The integral in Eq.~(\ref{eq:finite_volume_potential}) is infrared divergent, which can be seen by expanding the integrand 
at $4\pi^2\rho/L^2 \gg 1$:
\beq
\prod_{i=1}^3\theta_3\left(-\frac{\pi \vec{r}_i}{L},e^{-4\pi^2 \rho/L^2}\right)\approx 1 + 2\left(\cos\frac{2\pi x}{L}  + \cos\frac{2\pi y}{L} + \cos\frac{2\pi z}{L} \right)e^{-4\pi^2 \rho/L^2}.
\eeq
The constant term can be traced back to the zero modes in the first line of Eq.~\eqref{eq:finite_volume_potential}. 
This divergence is spurious and absent in the continuum where the integrand $d^3 q/\vec{q}^2$ is finite as $\vec{q} \to \vec{0}$.
It can be removed by subtracting the zero mode contribution~\cite{Hasenfratz:1989pk},
\begin{align}
V_R(\vec{r}) & = -e^2\int_0^\infty d\rho\frac{1}{L^3}\left[
\prod_{i=1}^3\theta_3\left(-\frac{\pi \vec{r}_i}{L},e^{-4\pi^2 \rho/L^2}\right) - 1\right] \\
& \approx -\frac{e^2}{4\pi r}\left(1 - 2.84 \frac{r}{L} + 2.09\frac{r^3}{L^3} + \dots\right)\,,
\label{eq:regularized_cart_disc}
\end{align}
where in the last line we chose $\vec{r} = r \hat z$ and expanded in $r/L$.\footnote{Note that 
because the $\vec{n}$ lattice is not rotationally invariant, this choice of $\vec{r}$ is 
not fully generic. Since we do not know how the real depletion of states should be realized, we neglect angular dependence and merely use the finite volume technique to extract the $|\vec{r}|/L$ scaling of the finite-$L$ corrections.}
Even terms are absent in parentheses because $\theta_3$ is an even function of its first argument, and 
the numerical result in the final line is obtained by fitting a polynomial to $V_R(\vec{r})/(-e^2/4\pi r)$. These coefficients can also be computed semi-analytically 
by splitting the $\rho$ integration into a region with $\rho L^2 \ll 1$ (where one can perform a small $\rho$ expansion but not an $r/L$ expansion), 
and another where one can expand in $r/L$; this calculation is presented in the Appendix~\ref{sec:appxA}. 
We checked these two approaches agree. For our purposes, all that is important is the powers of $r/L$. The $r\to 0$ limit of Eq.~\eqref{eq:regularized_cart_disc} also matches the results of Ref.~\cite{Hasenfratz:1989pk}. 

The leading correction to the potential in Eq.~(\ref{eq:regularized_cart_disc}) is independent of $r$ and thus 
does not contribute to energy level splittings. Therefore, from the absence of an $(r/L)^2$ term in~(\ref{eq:regularized_cart_disc}), naively we might conclude that the impact of a depleted density of 
states on  energy levels starts at $\mathcal{O}((r/L)^3)$. However, our calculation involved a sum over 
\emph{all} momenta, and it is not immediately clear whether these corrections are produced primarily by 
small  momenta, $q \ll 1/r$ (i.e. $n\ll  L/(2\pi r)$ modes), or momenta $q \sim 1/r$ (i.e. $n\sim  L/(2\pi r)$ modes).
As discussed in Sec.~\ref{sec:depletion_from_finite_volume} we are really interested in scale-dependent depletions, so 
we want to isolate the contributions of particular energy regimes to the $L$-dependence. We elect to focus on the highest energy scales that contribute significantly to the observable -- in the present case, the scales $q\sim 1/r$.

We  isolate the contribution of $q\sim 1/r$ momenta by considering the difference between the continuum and discrete calculation in 
different cells in $\vec{n} = L \vec{q}/(2\pi)$ space. We partition the continuum integration region in Eq.~\eqref{eq:qed_pot} into 
unit cubes corresponding to the terms in the discrete sum, such that the differences can be computed 
cell-by-cell. 
Since the integrand is even 
and the entire integration region is symmetric, we add the contributions from all $8$ cells that 
are related to each other by parity operations; the set of these 8 cells is denoted by $P_\vec{N}$.
For example, for the cell with the lowest vertex at $\vec{N}$ we find
\beq
4\pi^2 L\Delta_{\vec{N}} = \sum_{\vec{M}\in P_\vec{N}}\left(\int_{\vec{M}} d^3 n \frac{\cos (2\pi n_z r/L)}{\vec{n}^2} - \frac{\cos (2\pi M_z r/L)}{\vec{M}^2}\right),
\eeq
where the integration in the first term is over the cell. Details of this calculation are given in Appendix~\ref{sec:riemann_sum_error}. 

For $|\vec{N}|\gg 1$
the integrand is slowly-varying over the unit cube, so the integral can be performed by doing a Taylor expansion around $\vec{N}$. 
In the regime of interest $N_i \sim L/(2\pi r) \gg 1$, the result simplifies to 
\begin{align}
4\pi^2 L \Delta_{\vec{N}} &  \approx \frac{12}{N^4} \cos \left(\frac{2\pi N_z r}{L}\right)  \nonumber\\ 
& + \frac{4(2-N^2 (2\pi r/L)^2)}{3 N^4}\cos \left(\frac{2\pi N_z r}{L}\right) + \frac{16N_z (2\pi r/L)}{3N^4}\sin \left(\frac{2\pi N_z r}{L}\right),
\end{align}
where the first (second) line corresponds to the linear (quadratic) term in the Taylor expansion. 
There are $\mathcal{O}(N^3)$ such contributions, so the total finite-volume correction from $N_i  \sim L/(2\pi r)$ 
scales  as 
\beq
 \Delta_{\vec{N}} \times N^3 \sim  \left(\frac{1}{4\pi^2}\right)\left(\frac{r}{L^2} + \frac{r^2}{L^3} + \frac{r^3}{L^4} + \dots\right).
 \label{eq:cellscaling1}
 \eeq
Here we have dropped numerical coefficients and taken the trigonometric functions to be $\mathcal{O}(1)$. Eq.~(\ref{eq:cellscaling1}) is a somewhat surprising result, because the term of order $1/L^2$ 
 is \emph{absent} in the full result in Eq.~(\ref{eq:regularized_cart_disc}).\footnote{Note that the $r$-independent Casimir term in Eq.~\eqref{eq:regularized_cart_disc} is absent from Eq.~\eqref{eq:cellscaling1} because it is generated entirely by very low momenta, $q\sim 1/L$, i.e. $|\vec{n}|\sim 1$.} 
We have numerically confirmed this by  computing the differences between continuum and discrete contributions to the potential from momenta of order $1/r$.
We  conclude that the $1/L^2$ corrections cancel if the depletion of states (effective box size) is energy-\emph{independent}. 
If the depletion is instead energy-\emph{dependent}, this cancellation might be spoiled.
Since we do not know the precise functional form of the depletion, it is possible that the leading (additive) $r$-dependent corrections 
to the Coulomb potential scale as either $r/L^2$ or $r^2/L^3$. Below we will consider both possibilities in estimating the sensitivity of atomic 
level splittings to the depletion scale. 

\subsubsection{Finite Volume Effects in Hydrogen}
From the above, we are led to  consider corrections to the Coulomb potential  of the form
\beq
\Delta V(r) \sim \frac{e^2}{L} \left(\frac{r}{L}\right)^p,
\label{eq:finite_L_perturbation}
\eeq
where $p$ may be 1 or 2. The hydrogen wavefunctions are
\beq
\psi_{n\ell m} = \frac{c_{n \ell}}{a^{3/2}} \left(\frac{2r}{n a}\right)^\ell L_{n-\ell -1}^{2\ell+1}\left(\frac{2r}{n a}\right) e^{-r/(n a)} Y^m_\ell (\theta,\phi),
\eeq
where $a = 1/(\alpha m_e)$, $L^{q}_{p}(x)$ is an associated Laguerre polynomial and $c_{n\ell}$ is a normalization constant. 
The shifts to specific energy levels $\delta E_{n\ell}$ can be estimated as
\beq
\delta E_{n\ell} = \langle n\ell m |\Delta V| n \ell m \rangle = c_{n \ell}^2 \left(\frac{n}{2}\right)^3
\int_0^\infty du\, u^{2+2\ell}  \Delta V(a n u/2)  \left[L_{n-\ell -1}^{2\ell+1}(u)\right]^2 e^{-u}.
\label{eq:energy_shift}
\eeq
One of the most precisely-measured splittings measured is the  Lamb shift $\Delta E = E(2S_{1/2}) - E(2P_{1/2})$. 
The experimental value is~\cite{Lundeen:1981zz} 
\beq
\Delta E = 1057845(9) \;\mathrm{kHz},
\label{eq:lamb_shift_exp}
\eeq
in agreement with theoretical predictions (these depend on the proton radius so the two inconsistent values give 
somewhat different predictions~\cite{Eides:2007exa}). While there are slightly more precise experimental extractions of 
$\Delta E$ than Ref.~\cite{Lundeen:1981zz}, they rely on combinations multiple energy levels, complicating comparison with theory.
For a direct measurement of $\Delta E$ like~\cite{Lundeen:1981zz} the magnitude of the finite-$L$ correction is simply 
\beq
\delta(\Delta E) = |\delta E_{20} - \delta E_{21}| \sim \frac{e^2}{L} \left(\frac{a}{L}\right)^p.
\eeq
Demanding that this shift is less than the experimental uncertainty gives a lower bound on $L$:
\beq
\Leff(\mathrm{keV}) \gtrsim \begin{cases}
  10^{-3}\,\mathrm{m}  & p = 1 \\
  10^{-6}\,\mathrm{m} & p = 2,
\end{cases}
\eeq
where the characteristic energy is $\epsilon_* \sim \alpha m_e \sim \mathrm{keV}$. For comparison, 
the effective box size implied by the CKN DOS at this energy scale is 
\beq
L_{\mathrm{CKN}}(\mathrm{keV}) \sim 10^{14}\,\mathrm{m}.
\eeq

\subsection{Anomalous Magnetic Moment of Leptons}
\label{sec:g_minus_2}
The electron anomalous magnetic moment is one of the most precisely measured quantities in particle physics. Previous works have computed   the corrections to lepton $g-2$ in a finite volume~\cite{Hasenfratz:1989pk,Davoudi:2018qpl,Banks:2019arz,Cohen:2021zzr}. Since we are interested in energy-dependent effective volumes $\Leff(\epsilon)$, as above, we will need to modify these computations to isolate just the finite-volume terms that arise from a particular  energy scale of the individual particles in the intermediate state.

The ordinary one-loop infinite volume contribution to $g-2$ is~\cite{Schwinger:1948iu} 
\begin{align}
a=\frac{\alpha}{2\pi}.
\end{align}
We can rewrite the usual covariant Feynman integral in two useful ways. First, as a $3$-momentum integral, 
\begin{align}
a=6\pi\alpha \int \frac{d^3k}{(2\pi)^3}\frac{2k(k-\sqrt{k^2+m_e^2})+m_e^2}{3k^2m_e^2\sqrt{k^2+m_e^2}}.
\label{eq:F20k}
\end{align}
Here $k$ is the $3$-momentum of an internal electron. (All results in this section can be repeated for the muon with the replacement $m_e\rightarrow m_\mu$.) The external electrons are at rest and the external photon momentum is taken to zero. The integrand is dominated by $|k|\lesssim m_e$.  In old-fashioned perturbation theory, where the intermediate states are on-shell, the intermediate photon energy is of order $|k|$ and the electron energies are of order $\sqrt{k^2+m_e^2}$. Therefore, when $|k|\sim m_e$, all of the intermediate state particles have energies of order $m_e$. Thus we can estimate a lower bound on the DOS at energies of order $m_e$, $g(m_e)$, by isolating the contribution to $g-2$ from $|k|\sim m_e$ and computing the  effects of finite volume on it.

The second useful representation of the one-loop $g-2$ is the Schwinger parametrization, 
\begin{align}
a&=\frac{2\sqrt{\pi}\alpha}{ m_e^2}\int_0^\infty \frac{d\rho}{\sqrt{\rho}}\int\frac{d^3k}{(2\pi)^3}f(\rho)e^{-\rho k^2}\nonumber\\
&=\frac{\alpha}{4\pi m_e^2}\int_0^\infty \frac{d\rho}{\rho^2}f(\rho).
\label{eq:F20rho}
\end{align}
Here we have defined for convenience
\begin{align}
f(\rho)&\equiv 4m_e^4\rho^2\int_0^1 dz\, z (1-z)^2 e^{-\rho m_e^2(1-z)^2}\nonumber\\
&=-2+m_e\sqrt{\pi\rho}\,{\rm Erf}(m_e\sqrt{\rho})+2e^{-m_e^2\rho}.
\end{align}

Now we compute the sensitivity of $a$ to the density of states $g(\epsilon)$ by placing the process in a finite box of size $L^3$.
We discretize the momenta, $k\rightarrow 2\pi n/L$, and perform the sum over the $\vec n$ lattice. In the Schwinger parametrization the result can be expressed in terms of theta functions,
\begin{align}
\sum_{n=-\infty}^{\infty} e^{-\rho(2\pi n/L)^2} = \theta_3\left(e^{-4\pi^2\rho/L^2}\right) = \frac{L}{2 \sqrt{\pi\rho}}  \theta_3\left(e^{-L^2/4 \rho}\right)
\end{align}
where the two  expressions are related by Poisson summation and $\theta_3(\tau) = \theta_3(0,\tau)$ in the definition of Eq.~\eqref{eq:elliptic_theta_def}. 
In a finite box the zero mode does not contribute to Feynman sums and must be omitted to avoid a spurious infrared divergence in the integral over $\rho$. Including this subtraction, we have 
\begin{align}
a^L&\equiv\frac{2\sqrt{\pi}\alpha}{ m_e^2}\int_0^\infty \frac{d\rho}{\sqrt{\rho}} f(\rho) L^{-3}\left(\theta_3\left(e^{-\frac{4\pi^2\rho}{L^2}}\right)^3-1\right)\nonumber\\
&=\frac{\alpha}{4\pi m_e^2}\int_0^\infty \frac{d\rho}{\rho^2}f(\rho)\left(\theta_3\left(e^{-\frac{L^2}{4\rho}}\right)^3-\frac{8\pi^\frac{3}{2}\rho^\frac{3}{2}}{L^3}\right).
\label{eq:F20rhoL}
\end{align}

We can obtain an approximation to Eq.~(\ref{eq:F20rhoL}) as a series in $1/(m_e L)\ll 1$. Details are given in Appendix~\ref{sec:appxA}; here we give only the result:
\begin{align}
a_L\approx \frac{\alpha}{2\pi}\left( 1-8.91 (m_e L)^{-1} +35.65  (m_e L)^{-2} -59.21  (m_e L)^{-3} \right).
\label{eq:afiniteL}
\end{align}
The leading correction, $-8.91  (m_e L)^{-1}$, is similar in magnitude to the result obtained by Hasenfratz and Leutwyler~\cite{Hasenfratz:1989pk} for the first corrections to the two-point function; the integrand in that case is equivalent to ours in the large-$\rho$ limit of $f(\rho)$.\footnote{$f(\rho)\sim -2+\sqrt{\pi\rho m_e^2}$ at large $\rho$, up to exponentially small corrections. The appearance of $\sqrt{m_e^2}$ in this asymptotic expansion is responsible for the odd powers of $m_e$ appearing in Eq.~(\ref{eq:afiniteL}), despite the fact that the integrand in Eq.~(\ref{eq:F20rhoL}) is an even function of $m_e$. }

The finite $L$ effects on the entire mode sum are of order $L^{-1}$. However, our interest is in the contribution to the finite $L$ effects from modes around $k\sim m_e$. In fact, it is easy to see that the $1/L$ contribution in~Eq.~(\ref{eq:afiniteL}) comes entirely from the very low-momentum part of the mode sum, where $k\sim 1/L$. Finite $L$ effects are equivalent to the error introduced by performing a Riemann sum $\vec k\rightarrow 2\pi \vec n /L$ instead of an integral over $k$. 
The error contribution from the momentum bin at $\vec{N}$ (summed over all 8 bins related by parity) is
\beq
\Delta_{\vec{N}} = \frac{6\pi\alpha}{L^3} \sum_{\vec{M}\in P_\vec{N}}\left[ 
  \int_\vec{M} d^3 n \,g\left(\frac{2\pi n}{m_e L}\right) - g\left(\frac{2\pi M}{m_e L}\right)
\right],
\eeq
where $g(k/m_e)$ is the dimensionless form of the integrand in Eq.~\eqref{eq:F20k}.
Repeating the Riemann error analysis of Sec.~\ref{sec:coulomb} yields a total error estimate from the $\mathcal{O}(N^3)$ cells with $N \sim L m_e/(2\pi) \gg 1 $ (corresponding to $k\sim m_e)$ 
\beq
\Delta_N \times N^3 \sim \frac{6\pi \alpha}{(m_e L)^2}.
\label{eq:gm2_riemann_error}
\eeq
We conclude that the $1/L$ term in Eq.~(\ref{eq:afiniteL}) arises from {\emph{low energy}}  modes, rather than typical modes of $k\sim m_e$. On the other hand, generic contributions from $k\sim m_e$ are expected to be $\calO(1/L^2)$ by the  Riemann sum argument above. 
As before, the summation of all eight cells related by parity is key in obtaining the contribution of $k \sim m_e$ modes.
These conclusions are readily verified numerically, performing sums over low and high momentum modes and comparing to integration over the same $k$-volumes. 
Thus, we will use the term of order $1/L^2$ in Eq.~(\ref{eq:afiniteL}) to estimate the effect on $g-2$ of depleting the density of states near $k\sim m_e$. This is in contrast to Ref.~\cite{Banks:2019arz}, which used $1/L$ scaling; as we have just argued, 
however, such corrections are associated with low energies rather than $k\sim m_e$.

We are now in a position to convert these corrections into limits on the single particle DOS at $\epsilon \sim m_e$, parametrized by an effective box size $\Leff(m_e)$, and we can do the same for the muon. For our purposes we will neglect existing discrepancies between experiment and theory. $a_e$ has been measured to about $1\,$ppb and $a_\mu$ to about $0.3\,$ppm. Using the $1/L^2$ scaling for the corrections we find
\begin{align}
\Leff(m_e)&\gtrsim 10^5 m_e^{-1}\simeq 10\,{\rm nm}\nonumber\\
\Leff(m_\mu)&\gtrsim 10^4 m_\mu^{-1}\simeq 100\,{\rm fm}.
\end{align}
Note that we would obtain a much stronger bound if we discretized the state space by $1/L$ at all scales; in this case, $a$ is corrected at $\calO(1/L)$. 
This is the standard scaling for finite volume corrections to $g-2$ in a fixed box~\cite{Davoudi:2014qua,Banks:2019arz} and was used both in~\cite{Banks:2019arz} and a more recent analysis of the standard correlated UV-IR cutoff interpreation of CKN~\cite{Cohen:2021zzr}.
We are purposefully not using this scaling: as we have seen, most of the effects from a real finite box come from the lowest momenta of intermediate states, where the intermediate photon  is much softer than $m_e$. We elect to focus on the highest energy scale accessible to $g-2$, $m_e$, for two reasons: first, this is the scale where all the intermediate state particles are (semi)-relativistic and have energies of similar order; and second, the DOS at higher energies is of greater interest for the cosmological constant problem. 

The effective sizes implied by CKN scaling are
\begin{align}
L_{\rm CKN}(m_e)&\simeq 10^{21} m_e^{-1}\simeq 10^5\,{\rm km}\nonumber\\
L_{\rm CKN}(m_\mu)&\simeq 10^{19} m_\mu^{-1}\simeq 20\,{\rm km} .
\end{align}
Clearly we are in no danger of testing these densities.

\subsection{Higher Energies}
The finite volume technique used above for the Lamb shift and $g-2$ could be applied to other precision  processes. However, at much higher energies observables are both predicted and measured with much lower precision. Instead we make some simpler qualitative estimates.

Many detectors in high energy physics have energy resolutions of order $0.1-10\%$. For example, the electromagnetic calorimeter at CMS is able to measure electron momenta to order $1\%$ over a range 100 GeV - 1 TeV~\cite{CMS:2018ipm}. Astrophysical observatories have access to even higher energies. For example, IceCube observes PeV neutrinos with an energy resolution of $\sim 20\%$~\cite{IceCube:2013ccs}, while the Pierre Auger Observatory measures $\sim 100$ EeV cosmic rays with 
a $7\%$ resolution~\cite{PierreAuger:2020qqz}. This implies an empirical bound on the effective length, $\Leff({\rm 100\; GeV})\gtrsim 1\;{\rm fm}$, 
$\Leff({\rm 10^6\; GeV})\gtrsim 10^{-5}\;{\rm fm}$ and $\Leff({\rm 10^{11}\; GeV})\gtrsim 10^{-10}\;{\rm fm}$. These length scales are a factor 
of $10^{13}$, $10^9$ and $10^5$ smaller than implied by the CKN-inspired depletion of DOS in Eq.~\eqref{eq:naiveDDOS}.

\section{Hot Boxes and Depleted DOS}
\label{sec:hot_box}
Now we return to the CKN bound, examining the relation between hot box thought experiments and the depleted DOS of Eq.~(\ref{eq:naiveDDOS}) in more detail.
We begin by generalizing  Eq.~(\ref{eq:naiveDDOS}) to a family of modified  single particle relativistic DOS,
\begin{align}
g(\epsilon) = \epsilon^2 \Leff(\epsilon)^3,\;\;\; \Leff(\epsilon) = \min\left(L,M_p^{-1}(M_p/\epsilon)^n\right)
\label{eq:modified_dos}
\end{align}
for some $n>1$. We ignore any ${\cal{O}}(1)$ numbers. 
 It is also convenient to introduce the transition scale
\begin{align}
\hat \epsilon \equiv M_p/(M_pL)^{1/n}.
\end{align}
We take $n>1$ so that the transition energy satisfies $\hat\epsilon\gg 1/L$ for all systems larger than the Planck length.
For single particle energies $\epsilon < \hat \epsilon $, the density of states is depleted by  ordinary finite volume effects related to the physical size of the system under consideration. For 
$\epsilon > \hat \epsilon$ it is depleted more strongly. We could further generalize the factors of $M_p$ in Eq.~(\ref{eq:modified_dos}) to some lower scale associated with quantum gravity, if such a scale exists, but for simplicity we will keep it as $M_p$. 

Thus we have a one-parameter family of models labeled by the depletion rate parameter $n$. For large $n$, the depletion rate as a function of energy is  rapid, but the transition scale is also  high. For small $n$, the depletion rate with energy is slow, but the transition scale is low. Primarily we will  be interested in $n\sim 1-2$, as we will see below.

Note that since in any experimental system
$L M_p \gg 1$, the transition between standard and depleted density occurs at scales far below $\Mpl$.
We will also assume a  cutoff on QFT at $E\sim \Mpl$.

Now we estimate the effects of the DOS depletion on  hot box gravitational backreaction, and the radiative correction to the cosmological constant, as a function of the depletion rate parameter $n$ in Eq.~(\ref{eq:modified_dos}).

\subsection{Hot Boxes}

The energy and entropy of a thermal state of free particles are
\begin{align}
  E &=\int d \epsilon\, \epsilon \,g(\epsilon) f_{\pm}(\beta \epsilon)\nonumber\\
  S &=\int d \epsilon\,  g(\epsilon) \left[\beta \epsilon f_{\pm}(\beta \epsilon) \pm \log\left(1\pm e^{-\beta \epsilon}\right)\right]\nonumber\\
  &= -\int d \epsilon\,  g(\epsilon) \left[f_{\pm} \ln f_{\pm} \pm (1\mp f_{\pm})\ln\left(1\mp f_{ \pm}\right)\right]
\end{align}
where $f_{\pm}$ is the Fermi-Dirac (+) or Bose-Einstein (-)   distribution function and $\beta=1/T$. We will estimate the energy and the entropy for the depleted density of states in Eq.~(\ref{eq:modified_dos}).

First we consider the Fermi-Dirac case.
For high but subplanckian temperatures $M_p \gg T\gg \hat \epsilon$ the energy scales as 
\begin{align}
E & = \int^{\hat \epsilon} d\epsilon\, \frac{\epsilon^3 L^3}{e^{\beta \epsilon} + 1} + \int_{\hat \epsilon} d\epsilon \,\frac{(\Mpl/\epsilon)^{3n-3}}{e^{\beta \epsilon} + 1} \nonumber\\ 
& \sim \Mpl 
\begin{cases} 
  (T/\Mpl)^{4-3n} & 1< n < 4/3 \\
  \ln \left[(L T)/(L \Mpl)^{1/4}\right]                 & n = 4/3 \\ 
(L \Mpl)^{3-4/n}  & n > 4/3,
\end{cases}
\end{align}
where we drop $\mathcal{O}(1)$ factors and take $n>1$ for reasons described above. For $n>4/3$, the energy is dominated by contributions from modes near $\hat \epsilon$. As in CKN we can compare the Schwarzschild scale $E/M_p^2$ to the system size $L$. The bound $L>E/M_p^2$ is satisfied for  $n<2$ and saturated for $n=2$.
For the entropy, the scaling behavior is
\begin{align}
S \sim (L M_p)^{3-3/n}.
\end{align}

In the Bose-Einstein case, at high $ M_p \gg T \gg \hat \epsilon$, the energy behaves as 
\begin{align}
E & = \int^{\hat \epsilon} d\epsilon\, \frac{\epsilon^3 L^3}{e^{\beta \epsilon} - 1} + \int_{\hat\epsilon} d\epsilon\, \frac{(\Mpl/\epsilon)^{3n-3}}{e^{\beta \epsilon} - 1}  \sim T (L \Mpl)^{3-3/n}.
\end{align}
again dropping $\mathcal{O}(1)$ factors. This scaling applies for $n>1$, and is different 
from the Fermi-Dirac case due to large occupation numbers in low energy modes. Comparing the Schwarzschild scale $E/M_p^2$ to the system size $L$,  we find that black holes are not formed for any  $T < \Mpl$ if $n\leq 3/2$. 
For $n > 3/2$, however,  there is still a bound on the temperatures,
\beq
T < \frac{\Mpl}{(L\Mpl)^{2-3/n}}.
\eeq
In fact, for $n=2$ the bound on $T$ obtained in this manner is approximately the same as in the original CKN bound. So in this case, the depletion of the density of states does not automatically eliminate black holes at high temperatures in finite-size bosonic systems. 

On the other hand, {\emph{no}} depleted single-particle DOS {\emph{or}} correlated UV-IR cutoff on Feynman integrals can remove black holes formed by non-thermal states of soft bosonic modes with enormous occupation numbers. Some states must be removed in a way that depends on the occupancy. Instead, however, we can ask whether the entropy is reduced in a way consistent with the CKN bound. The entropy  is 
\begin{align}
S\sim (L M_p)^{3-3/n}\log(\beta \hat \epsilon).
\end{align}
Up to the logarithm, it saturates the $L^{3/2}$  scaling for $n=2$.

From these thought experiments, we conclude that the range $1<n\leq 2$ is of interest, with possibly special roles played by $n=2$ and  $n=3/2$. $n=2$, the model used for comparison in Sec.~\ref{sec:depletion_from_finite_volume}, is the more conservative of the two: it corresponds a faster rate of depletion, but it turns on at higher energies, corresponding to a lesser overall depletion at any given scale. 

\subsection{Cosmological Constant}
Finally let us consider consider the impact of a depleted DOS on the cosmological constant. 
In the power-law models above, the QFT vacuum energy contribution to the cc is finite if $n > 4/3$ and scales as
\begin{align}
  \delta\rho&\sim H^{3}\int d \epsilon\, \epsilon \,g(\epsilon) \sim \hat\epsilon^4
  \label{eq:cc_correction}
\end{align}
where  $H$ is the Hubble scale. Unlike the local observables considered in Sec.~\ref{sec:depletion_from_finite_volume}, 
the cc does not have a preferred energy scale $\epsilon_*$; rather, $\delta\rho$ is dominated by the most numerous modes 
around $\hat \epsilon$.

The relative size of the correction in Eq.~\eqref{eq:cc_correction} is 
\beq
\frac{\delta \rho}{\rho} \sim \left(\frac{\rho}{\mpl^4}\right)^{2/n -1}.
\eeq
We  see that for $n\leq 2$, the cc appears to be technically natural, i.e. the quantum correction is at most of the size of the 
cc itself. While $n>2$ still gives rise to a finite $\delta \rho$, the correction exceeds the observed cc because the 
transition scale $\hat\epsilon$ becomes too large.
For $n = 2$, $\hat \epsilon \sim \mathrm{meV}$. We emphasize that we do not interpret this as a transition 
energy beyond which QFT breaks down in all experiments. Rather, $\epsilon > \hat \epsilon$ merely marks the energy scale at which the characteristic momentum spacing between independent field modes exceeds $H\sim 10^{-33}$ eV. 
Systems much smaller than the horizon size (i.e., laboratory experiments) are sensitive to a different DOS (i.e., different $L$ and $\hat\epsilon$) 
and therefore cannot constrain fluctuations over $\sim 1/H$, as emphasized by CKN~\cite{Cohen:1998zx}. 
This point of view is in contrast with the idea that  
laboratory experiments which measure the gravitation of vacuum fluctuations can be used to justify the standard (UV-sensitive) QFT estimate of the vacuum energy~\cite{Polchinski:2006gy}. In fact, the $(g-2)$ calculation in Sec.~\ref{sec:g_minus_2} shows that even some of the most precise probes of vacuum fluctuations are not sensitive to the depletion of the DOS needed to address the cc fine-tuning problem.

\section{Conclusion}
\label{sec:conclusion}

Quantum field theory provides an accurate description of local observables over an enormous range of energy scales, including high energy collider processes and precision measurements in atomic physics. However, a naive application of QFT to extensive systems runs 
into conceptual problems, including the violation of entropy bounds or strong gravitational backreaction even for rather 
mundane energy densities and volumes. For example, a $10$ m room filled with $T\sim 10$ GeV plasma is within its own Schwarzschild radius, implying 
that nongravitational QFT must break down. 
Based on these thought experiments, CKN proposed a correlation between the UV 
and the IR cutoffs on QFT~\cite{Cohen:1998zx}. 
We have explored an alternative interpretation proposed in~\cite{Banks:2019arz} in which the 
independent degrees of freedom are depleted in an energy scale-dependent way. 
An analogous depletion has been suggested  in partonic models of holography~\cite{Susskind:1994vu}. 
We studied a simple phenomenological implementation of a state depletion, motivated by the ``hot box" thought experiments of CKN.

For local observables, the DOS depletion can be viewed as a finite volume effect, where the volume depends on the characteristic 
energy scale of the observable. We have applied this reasoning to estimate the corrections to two sensitive probes of QFT, lepton $g-2$ and the hydrogen Lamb shift, finding that despite their incredible precision, these measurements are far from being sensitive to the depletions motivated by quantum gravity. Tests of the DOS are, however, a novel axis for testing QFT. It would be interesting explore other observations that probe the DOS at different energy scales, such as high energy cosmic ray scattering. 

An interesting application of these DOS models is the calculation of the QFT vacuum energy contribution to the cosmological constant,
which becomes UV-insensitive if the depletion is rapid enough. For a DOS that saturates the CKN bound, we find that 
the cc is technically natural. 

The most important questions, unaddressed here, are how Lorentz invariance emerges and whether the QFT entropy scaling $S_{QFT}\sim(M_p L)^{3/2}$ can be motivated by microscopic models analogous to the partonic models of holography. The Holographic Space Time models of Banks and Fischler~\cite{Banks:2013fr,Banks:2018aed,Banks:2020zcr}, for example, have been argued to realize just such a scaling~\cite{tbpc}.

\section*{Acknowledgements}
We thank Tom Banks, Ben Lillard, Aleksey Cherman and Theo Jacobson for useful discussions. PD acknowledges support from the US Department
of Energy under Grant No. DE-SC0015655. 
This manuscript has been authored by Fermi Research Alliance, LLC under Contract No. DE-AC02-07CH11359 with the U.S. Department of Energy, Office of Science, Office of High Energy Physics. 

\appendix
\section{Analytic Results for Finite Volume Corrections}
\label{sec:appxA}
In this Appendix we give details of the calculation of the $1/L$ series for the Coulomb problem and lepton magnetic moments, 
Eqs.~\eqref{eq:regularized_cart_disc} and~\eqref{eq:afiniteL}, respectively.

\subsection{Coulomb Potential}
We begin by splitting the $\rho$ integration in the first 
line of Eq.~\eqref{eq:regularized_cart_disc} into intervals $[0,\hat\rho]$ and $(\hat\rho,\infty)$, where $\hat\rho$ 
is chosen such that $4\pi^2 \hat\rho/L^2 \ll 1$. In the first interval, we can expand the integrand 
in $4\pi^2 \rho/L^2 \ll 1$, perform the integration and then expand the result in $r/L$:
\begin{align}
I_1 & = -e^2\int_0^{\hat\rho} d\rho\frac{1}{L^3}\left[
\theta_3\left(0,e^{-4\pi^2 \rho/L^2}\right)\theta_3\left(0,e^{-4\pi^2 \rho/L^2}\right)\theta_3\left(-\frac{r}{L},e^{-4\pi^2 \rho/L^2}\right) - 1\right]\\
&\approx -\frac{e^2}{4\pi r} \left[1 - \frac{r}{L}\left(4\pi \hat\rho/L^2 + L/\sqrt{\pi \hat\rho}\right) + \frac{r^3}{12 \sqrt{\pi}\hat\rho^{3/2}}\right] 
\end{align}
The final expansion in $r/L$ is valid if $r/L \ll \sqrt{\hat\rho/L^2}$.
In the second interval, $(\hat\rho,\infty)$, the integrand can be expanded in $r/L$ because the function is analytic once the singularity at $\rho = 0$ is excluded. 
The result is
\begin{align}
  I_2 & = -e^2\int_{\hat\rho}^\infty d\rho\frac{1}{L^3}\left[
\theta_3\left(0,e^{-4\pi^2 \rho/L^2}\right)\theta_3\left(0,e^{-4\pi^2 \rho/L^2}\right)\theta_3\left(-\frac{r}{L},e^{-4\pi^2 \rho/L^2}\right) - 1\right]\\
& \approx -\frac{e^2}{4\pi r} \sum_{k=0} c_{2k + 1} (r/L)^{2k+1},
\end{align}
where the first two coefficients are given by
\begin{subequations}
  \begin{align}
    c_1 & = \int_{\hat\rho}^\infty d\rho \left[\theta_3\left(0,e^{-4\pi^2 \rho/L^2}\right)^3 - 1\right]\\
    c_3 & = \int_{\hat\rho}^\infty d\rho \frac{\pi^2}{2}\theta_3\left(0,e^{-4\pi^2 \rho/L^2}\right)^2\theta_3^\prime \left(0,e^{-4\pi^2 \rho/L^2}\right).
  \end{align}
\end{subequations}
While $I_1$ and $I_2$ individually depend on the unphysical parameter $\hat\rho$, their sum does not, which we checked for a large 
range of $\hat\rho /L^2$. The resulting $r/L$ expansion of $V_R = I_1 + I_2$ agrees with the one given in Eq.~\eqref{eq:regularized_cart_disc} 
obtained by a fully-numerical method.

\subsection{Anomalous Magnetic Moment}
In this appendix we derive the $1/L$ expansion for the finite-volume corrections to $g-2$ given in Eq.~(\ref{eq:afiniteL}).
To simplify notation, we define
\begin{align}
y\equiv m_e L.
\end{align}
Let $t=\frac{L^2}{4\pi \rho}$. In terms of $t$, we have
\begin{align}
a_L = \frac{\alpha}{y^2}\int_0^\infty dt\, t^{-3/2} f\left(\frac{L^2}{4\pi t}\right)\left(\theta_3\left(e^{-\pi/t}\right)^3-1\right)
\end{align}
Now we split the integration into two regimes: $0\leq t \leq y$, and $y < t < \infty$. In these regimes different parts of the integrands can be replaced by asymptotic expansions, up to terms exponentially small in $y$:
\begin{align}
a_L &= \frac{\alpha}{2\pi}\left( \calI_1 + \calI_2\right)\nonumber\\
\calI_1&\approx2\pi\int_0^{y} dt\, \left(-\frac{2}{y^2t^{\frac{3}{2}}}+\frac{1}{2yt^2}\right)\left(\theta_3\left(e^{-\pi/t}\right)^3-1\right)\nonumber\\
\calI_2&\approx 2\pi\int_{y}^\infty dt\, \left(-\frac{2}{y^2t^{\frac{3}{2}}}+\frac{2}{y^2t^{\frac{3}{2}}}e^{-\frac{y^2}{4\pi t}}+\frac{1}{2yt^2}{\rm Erf}\left(\frac{y}{2\sqrt{\pi t}}\right)\right)\left(t^{3/2}-1\right).
\end{align}
First consider $\calI_1$. We can further split the integration range into $0\leq t\leq \hat t$ and $\hat t< t<y$, for some $1\ll\hat t\ll y$. $\hat t$ is arbitrary as long as it falls in this range. Then we can write
\begin{align}
\calI_1 &= \frac{c_1(\hat t)}{y} + \frac{c_2(\hat t)}{y^2} +2\pi\int_{\hat t}^ydt\,\left(-\frac{2}{y^2t^{\frac{3}{2}}}+\frac{1}{2yt^2}\right)\left(t^\frac{3}{2}-1\right)\nonumber\\
c_1 &= 2\pi \int_0^{\hat t}dt\,\left(\frac{1}{2t^2}\right)\left(\theta_3\left(e^{-\pi/t}\right)^3-1\right)\nonumber\\
c_2 &= 2\pi \int_0^{\hat t}dt\,\left(-\frac{2}{t^{\frac{3}{2}}}\right)\left(\theta_3\left(e^{-\pi/t}\right)^3-1\right),
\end{align}
where $c_1$ and $c_2$ can be evaluated numerically for some arbitrary choice of $\hat t$, and the remaining integral can be evaluated analytically. We find:
\begin{align}
\calI_1 = 2\pi\left(y^{-1/2} -3.42 y^{-1} +6.17 y^{-2} -4 y^{-5/2}\right).
\end{align}
$\calI_2$ can be treated similarly. It is convenient to redefine $\tilde t = t/ y^2 $ so that
\begin{align}
\calI_2 = 2\pi \int_{1/y}^\infty d\tilde t\,\left(-2+2e^{-\frac{1}{4\pi \tilde t}}+\frac{1}{2\sqrt{\tilde t}}{\rm Erf}\left(\frac{1}{2\sqrt{\pi \tilde t}}\right)\right)
\left(1-\frac{1}{\tilde t^{3/2}y^3}\right),
\end{align}
and further split the integration range into $1/y\leq \tilde t\leq \hat {\tilde t}$ and $\hat {\tilde t}< \tilde t<\infty$, for some $1/y\ll\hat {\tilde t}\ll 1$. As before, $\hat {\tilde t}$ is arbitrary as long as it falls in this range. Then 
\begin{align}
\calI_2 &= c_0(\hat {\tilde t}) + \frac{c_3(\hat {\tilde t})}{y^3} +2\pi\int_{1/y}^{\hat {\tilde t}} d\tilde t\,\left(-2+\frac{1}{2\sqrt{\tilde t}}\right)
\left(1-\frac{1}{\tilde t^{3/2}y^3}\right)\nonumber\\
c_0 &= 2\pi\int_{\hat {\tilde t}}^\infty d\tilde t\,\left(-2+2e^{-\frac{1}{4\pi \tilde t}}+\frac{1}{2\sqrt{\tilde t}}{\rm Erf}\left(\frac{1}{2\sqrt{\pi \tilde t}}\right)\right)\nonumber\\
c_3 &= 2\pi\int_{\hat {\tilde t}}^\infty d\tilde t\,\left(-2+2e^{-\frac{1}{4\pi \tilde t}}+\frac{1}{2\sqrt{\tilde t}}{\rm Erf}\left(\frac{1}{2\sqrt{\pi \tilde t}}\right)\right)
\left(-\frac{1}{\tilde t^{3/2}}\right),
\end{align}
where, as before, $c_0$ and $c_3$ can be evaluated numerically for some arbitrary choice of $\hat {\tilde t}$, and the integral can be evaluated analytically. We find:
\begin{align}
\calI_2 = 2\pi\left(0.159-y^{-1/2} +2 y^{-1} +\frac{1}{2} y^{-2} +4 y^{-5/2}-9.42y^{-3}\right).
\end{align}
All the fractional powers of $y$ cancel between $I_1$ and $I_2$. In total, and restoring $y\rightarrow m_e L$,
\begin{align}
a_L\approx \frac{\alpha}{2\pi}\left( 1-8.91 (m_e L)^{-1} +35.65  (m_e L)^{-2} -59.21  (m_e L)^{-3} \right).
\end{align}

\section{Finite Volume Effects from Riemann Sum Error}
\label{sec:riemann_sum_error}
In this Appendix we  derive the scaling of finite-box corrections to the Coulomb potential as a function of the exchanged photon's energy scale by directly 
computing the difference between continuum and discretized versions in each momentum cell.

We define the complete continuum potential integral as
\beq
f(r) = \frac{1}{4\pi^2 L}\int d^3 n \frac{\cos (2\pi n_z r/L)}{\vec{n}^2},
\label{eq:pot_cont_int}
\eeq
where we changed variables to the dimensionless momentum $\vec{n} = \vec{q} L /(2\pi)$; 
we have not made use of the three parity symmetries yet, so each integral 
is over $(-\infty,+\infty)$.
The corresponding discrete sum is 
\beq
\hat{f}(r) = \frac{1}{4\pi^2 L} \sum_{\vec{N}}^\prime\frac{\cos (2\pi N_z r/L)}{\vec{N}^2},
\label{eq:pot_disc_int}
\eeq
where we used capital $N_i$ to distinguish the summation indices from the integration variables $n_i$; the prime 
denotes omission of $\vec{N}=\vec{0}$. 
We define the difference
\beq
\Delta = f(r) - \hat{f}(r). 
\eeq
This is most easily computed if we divide up the $d^3n$ integration region in Eq.~\eqref{eq:pot_cont_int} into 
unit cubes corresponding to each term in the sum Eq.~\eqref{eq:pot_disc_int}:
\beq
f(r) = \frac{1}{4\pi^2 L} \sum_{\vec{N}} \int_{\vec{N}} d^3 n \frac{\cos (2\pi n_z r/L)}{\vec{n}^2},
\label{eq:pot_cont_int_partitioned}
\eeq
where each cube has its lowest vertex at $\vec{N}$. 

There are 8 boxes/integration regions that have $\vec{0}$ at a vertex of their boundary, so we treat these separately; 
these 8 cubes have starting vertices at 
\begin{align}
  (-1,-1,-1),\; (-1,-1,0),\; (-1,0,-1),\; (0,-1,-1) \nonumber\\
  (0,0,0),\; (0,0,-1),\; (0,-1,0),\; (-1,0,0).
  \label{eq:boxes_near_origin}
\end{align}
Since in this 8 box region $n_z \leq 1$ and we are interested in $r/L \ll 1$ we can expand the cosine in the integrand and add up the 
numerically-evaluated contributions of all 8 boxes, yielding
\beq
4\pi^2 L \Delta_{|\vec{N}|\leq 1}  
\approx 10.5 - \frac{5}{2} \left(\frac{2\pi r}{L}\right)^2
\eeq
where $10.5$ and $5/2$ result from doing $d^3 n$ integrals numerically. The same argument holds 
for other regions with $|\vec{N}|\sim 1$ -- they contribute at 
$\mathcal{O}(1/L)$ and $\mathcal{O}(r^2/L^3)$ to $\Delta$, since there are $\mathcal{O}(1)$ such terms.

For $|\vec{N}|\gg 1$, the integrands in each term in $\Delta$ are slowly varying, so we can 
perform a Taylor expansion in each box, $\vec{N}$:
\beq
\frac{\cos (2\pi n_z r/L)}{\vec{n}^2} -\frac{\cos (2\pi N_z r/L)}{\vec{N}^2}
\approx \nabla_i g (\vec{n} - \vec{N})_i + \frac{1}{2} \nabla_i \nabla_j g(\vec{n} - \vec{N})_j (\vec{n} - \vec{N})_j + \dots
\label{eq:taylor_exp}
\eeq
where  $g$ is a shorthand for $\cos (2\pi n_z r/L)/\vec{n}^2$; the derivatives can be evaluated, while the $d^3 n$ give
\begin{align}
  \int_{n_x}^{n_x+1}\int_{n_y}^{n_y+1}\int_{n_y}^{n_y+1} d^3 n (\vec{n} - \vec{N}) & = \frac{1}{2}(1,1,1) \\
\int_{n_x}^{n_x+1}\int_{n_y}^{n_y+1}\int_{n_y}^{n_y+1} d^3 n (\vec{n} - \vec{N})\otimes (\vec{n} - \vec{N}) & =
\begin{pmatrix}
  1/3 & 1/4 & 1/4 \\
  1/4 & 1/3 & 1/4 \\
  1/4 & 1/4 & 1/3
\end{pmatrix}.
\end{align}
We must add all regions related by parity to exhibit the symmetries of the whole integration region and integrand. 
There are three parity transformations that can be used individually, or combined, such that 
a single region is related to seven others by these transformations; 
parity in the $i$'th direction maps $-N_i - 1 \to + N_i$.
Therefore if we start with a box with its lowest vertex at $(N_x, N_y, N_z)$ we relate it to 7 others by these parity 
operations. We denote the whole set of boxes generated in this way by $P_\vec{N}$:
\begin{align}
  P_{\vec{N}}  = &\Big\{(-N_x -1, -N_y -1, -N_z -1),\;(N_x, -N_y - 1, -N_z - 1),\nonumber\\
               &(N_x,  N_y, -N_z - 1),\; (N_x, N_y, N_z),\nonumber\\
               &(-N_x - 1, N_y, -N_z - 1),\; (-N_x - 1, N_y, N_z)\nonumber\\
             &(-N_x - 1, -N_y - 1, N_z),\; (N_x, -N_y - 1, N_z)\nonumber\Big\}.
\end{align}
For example, taking $\vec{N} = \vec{0}$ gives the set of 8 boxes 
surrounding the origin in Eq.~\eqref{eq:boxes_near_origin}. 
Adding up the contributions of all these boxes for $N \gg 1$ and $N_z \gg 1$ yields an error of 
 \beq
 4\pi^2 L \Delta_{N\gg 1,N_z\gg 1} \approx 
 \frac{12}{N^4}\cos N_z \epsilon 
 + \frac{4(2 - N^2 \epsilon^2)}{3N^4}\cos N_z \epsilon + \frac{16 N_z \epsilon}{3 N^4}\sin N_z \epsilon,
 \label{eq:coulomb_riemann_error_partial}
 \eeq
 where $\epsilon = 2\pi r /L$ and the first (second and third) term arises from the first (second) term in the Taylor expansion 
 in Eq.~\eqref{eq:taylor_exp}. 
 In the regime $N \sim N_z \gg 1$ there are $\mathcal{O}(N^3)$ boxes that contribute ($\sim 4\pi N^2 dN$ with $dN \sim \mathcal{O}(N)$);
 the total contribution to the Riemann sum error is therefore of order $N^3$ times Eq.~\eqref{eq:coulomb_riemann_error_partial}. 
 For $N_i \sim L/r \sim 1/\epsilon$, the trigonometric functions are $\mathcal{O}(1)$ and we obtain the 
schematic scaling of $1/L$ corrections given in Eq.~\eqref{eq:cellscaling1}.

An analogous calculation can be performed for the $g-2$ integral in Eq.~\eqref{eq:F20k}. The result is given in Eq.~\eqref{eq:gm2_riemann_error}.

\bibliographystyle{JHEP}
\bibliography{biblio}
\end{document}